\documentclass[superscriptaddress,aps,prb,longbibliography,twocolumn]{revtex4-1}

\usepackage{graphicx}
\usepackage[colorlinks=True,linkcolor=red,citecolor=blue,urlcolor=blue]{hyperref}
\usepackage[version=4]{mhchem}
\usepackage{siunitx}
\usepackage{multirow}
\usepackage{bm}
\usepackage{amssymb}
\usepackage{array}
\usepackage{makecell}
\usepackage{braket}

\newcounter{Methods}

\newcounter{SI}

\usepackage{etoolbox}
\makeatletter         
\newcounter{firstbib} 

\AtBeginEnvironment{thebibliography}{
	
}

\apptocmd{\thebibliography}{
	\setcounter{NAT@ctr}{\value{firstbib}} 
	
}{}{}  

\AtEndEnvironment{thebibliography}{
	\setcounter{firstbib}{\value{NAT@ctr}}
	
}
\makeatother

\begin{document}

\title{Anisotropic electrical and thermal magnetotransport in the magnetic semimetal GdPtBi}

\author{Clemens Schindler}
\email{clemens.schindler@cpfs.mpg.de}
\affiliation{Max Planck Institute for Chemical Physics of Solids, 01187 Dresden, Germany}

\author{Stanislaw Galeski}
\affiliation{Max Planck Institute for Chemical Physics of Solids, 01187 Dresden, Germany}

\author{Walter Schnelle}
\affiliation{Max Planck Institute for Chemical Physics of Solids, 01187 Dresden, Germany}

\author{Rafa{\l}  Wawrzy\'{n}czak}
\affiliation{Max Planck Institute for Chemical Physics of Solids, 01187 Dresden, Germany}

\author{Wajdi Abdel-Haq}
\affiliation{Max Planck Institute for Chemical Physics of Solids, 01187 Dresden, Germany}

\author{Satya N. Guin}
\affiliation{Max Planck Institute for Chemical Physics of Solids, 01187 Dresden, Germany}

\author{Johannes Kroder}
\affiliation{Max Planck Institute for Chemical Physics of Solids, 01187 Dresden, Germany}

\author{Nitesh Kumar}
\affiliation{Max Planck Institute for Chemical Physics of Solids, 01187 Dresden, Germany}

\author{Chenguang Fu}
\affiliation{Max Planck Institute for Chemical Physics of Solids, 01187 Dresden, Germany}

\author{Horst Borrmann}
\affiliation{Max Planck Institute for Chemical Physics of Solids, 01187 Dresden, Germany}

\author{Chandra Shekhar}
\affiliation{Max Planck Institute for Chemical Physics of Solids, 01187 Dresden, Germany}

\author{Claudia Felser}
\affiliation{Max Planck Institute for Chemical Physics of Solids, 01187 Dresden, Germany}

\author{Tobias Meng}
\affiliation{Institute of Theoretical Physics, Technische Universit{\"a}t Dresden, 01062 Dresden, Germany}

\author{Adolfo G. Grushin}
\affiliation{Univ. Grenoble Alpes, CNRS, Grenoble INP, Institut N\'eel, 38000, Grenoble, France}

\author{Yang Zhang}
\affiliation{Max Planck Institute for Chemical Physics of Solids, 01187 Dresden, Germany}

\author{Yan Sun}
\affiliation{Max Planck Institute for Chemical Physics of Solids, 01187 Dresden, Germany}

\author{Johannes Gooth}
\email{johannes.gooth@cpfs.mpg.de}
\affiliation{Max Planck Institute for Chemical Physics of Solids, 01187 Dresden, Germany}

\date{\today}

\begin{abstract}
The half-Heusler rare-earth intermetallic GdPtBi has recently gained attention due to peculiar magnetotransport phenomena that have been associated with the possible existence of Weyl fermions, thought to arise from the crossings of spin-split conduction and valence bands.
On the other hand, similar magnetotransport phenomena observed in other rare-earth intermetallics have often been attributed to the interaction of itinerant carriers with localized magnetic moments stemming from the $4f$-shell of the rare-earth element.
In order to address the origin of the magnetotransport phenomena in GdPtBi, we performed a comprehensive study of the magnetization, electrical and thermal magnetoresistivity on two single-crystalline GdPtBi samples.
In addition, we performed an analysis of the Fermi surface via Shubnikov-de Haas oscillations in one of the samples and compared the results to \emph{ab initio} band structure calculations.
Our findings indicate that the electrical and thermal magnetotransport in GdPtBi cannot be solely explained by Weyl physics and is strongly influenced by the interaction of both itinerant charge carriers and phonons with localized magnetic Gd-ions and possibly also paramagnetic impurities.
\end{abstract}
\maketitle
\section{Introduction}
Weyl fermions can be realized as low-energy quasiparticles in certain semimetals with topologically protected crossing points of two inverted electronic bands with linear dispersion\cite{Xu2015,Lu2015,Armitage18}.
They occur in pairs of independent nodes, separated in momentum space with opposite chirality - a quantum number defining the `handedness' of a quasiparticle's spin relative to its momentum.
Classically, the particle number of each chirality is separately conserved.
However, at the quantum level electromagnetic fields can violate the conservation of the particle number at individual nodes due to quantum fluctuations.
This phenomenon is known as the chiral anomaly\cite{Nielsen1983,Bertlmann2000}, physically interpreted as simultaneous production of particles of one chirality and anti-particles of the opposite chirality.
In the context of Weyl semimetals, the chiral anomaly is expected to induce a steady out-of-equilibrium flow of quasiparticles between the left- and right-handed nodes, leading to a reduction of electrical and thermal magnetoresistivity\cite{Son2013,Kharzeev2014,Lundgren2014,Sharma2016,Lucas2016} in magnetic fields $\bm{H}$ aligned with the electric field $\bm{E}$ or thermal gradient $\bm{\nabla} T$, respectively.
By now, signatures for the chiral anomaly-induced negative magnetoresistance have been reported for electrical measurements of a number of semimetals\cite{Kim2013,Huang2015,Xiong2015,Li2016,Li2015,Wang2016,Lv2017,Zhang2016,Gooth2017}.
One of the challenges for identifying the chiral anomaly is to exclude other mechanisms leading to a similar effect, such as inhomogeneous currents paths in high-mobility materials\cite{Reis2016,Arnold2016}.
Concerning signatures of the chiral anomaly in thermal transport, the main experimental challenge is the extremely low density of electronic states in Weyl semimetals, which renders thermal transport heavily dominated by phononic conduction.
Nonetheless, a negative magnetothermal resistivity has recently been observed in a Bi-Sb alloy, where the Fermi level has been tuned to be exactly located at the band crossing points\cite{Vu2019}.
A prominent example for the observation of the chiral anomaly in electrical and thermoelectrical magnetotransport is the semimetal GdPtBi.
GdPtBi is a member of the half-Heusler \emph{R}PtBi series, where \emph{R} denotes a rare-earth element.
Density functional theory (DFT) calculations for these compounds yield an electronic band structure similar to the binary semiconductor HgTe\cite{Chadov2010}, whereby touching of conduction and valence band occurs near the Fermi level.
Depending on the strength of the spin-splitting, the \emph{R}PtBi compounds can be driven from a topologically trivial gapped state to a band-inverted state with topologically protected crossing points, potentially hosting Dirac or Weyl fermions\cite{Chadov2010}.
Alongside the predictions concerning the single-electron bands, the \emph{R}PtBi series exhibits a variety of many-body phenomena ranging from magnetism\cite{Ueland2014,Mun2013,Mun2016,Canfield1991,Kozlova2005,Mueller2015}, superconductivity\cite{Kim2018,Tafti2013} and heavy fermion behavior\cite{Ueland2014,Canfield1991,Mun2013}.
Among the \emph{R}PtBi series, GdPtBi probably is the most studied compound and thought to be a zero-gap semimetal, which upon application of an external magnetic field exhibits a band inversion\cite{Hirschberger2016}.
There have been reports on a chiral anomaly-induced negative longitudinal magnetoresistivity\cite{Hirschberger2016,Shekhar2016} and anisotropic planar Hall resistivity\cite{Kumar2018} as well as anomalous features in the Hall\cite{Suzuki2016,Shekhar2016} and (magneto-)resistivity\cite{Hirschberger2016,Suzuki2016,Mun2016,Shekhar2016}.
Many studies\cite{Mueller2014,Kreyssig2011,Canfield1991,Sukhanov2019} have also been carried out on the magnetism of GdPtBi, exploring the antiferromagnetic ordering of the local moments stemming from Gd`s half-filled $4f$-shell with $L=0$ and $J=S=7/2$.
Furthermore, optical methods have been employed, confirming the presence of electronic bands with nearly linear dispersion\cite{Hutt2018}.
As stated in ref.~\onlinecite{Chadov2010}, the many-body correlations of materials with partially filled $f$-shells cannot be treated in the local-density approximation used for calculating the band structure in refs.~\onlinecite{Chadov2010,Hirschberger2016}.
Therefore it remains an open question whether interactions of the Gd $4f$-electrons with the itinerant carriers play a role in the  observed magnetotransport phenomena in GdPtBi, or whether the single-electron picture is a valid assumption, acting as a foundation for the Weyl fermion scenario.
In order to address this question, we performed a comprehensive study of the physical properties of two distinct GdPtBi samples.
In Section~\ref{sec:Methods}, the crystal growth and related issues as well as measurement techniques are described.
Section~\ref{sec:Characterization} is about peculiarities in the magnetization of the two samples, as well as basic characterization of the electrical and thermal transport.
In Section~\ref{sec:Bandstructure}, the Shubnikov-de Haas oscillations occuring in one of the samples are analyzed and results of our DFT results are shown.
The analysis of the electrical and thermal magnetotransport are given in Sections~\ref{sec:MR} and \ref{sec:MTR}, respectively.
\section{Methods}
\label{sec:Methods}
\begin{figure}[ht!]
	\centering
	\includegraphics[width=8.6cm]{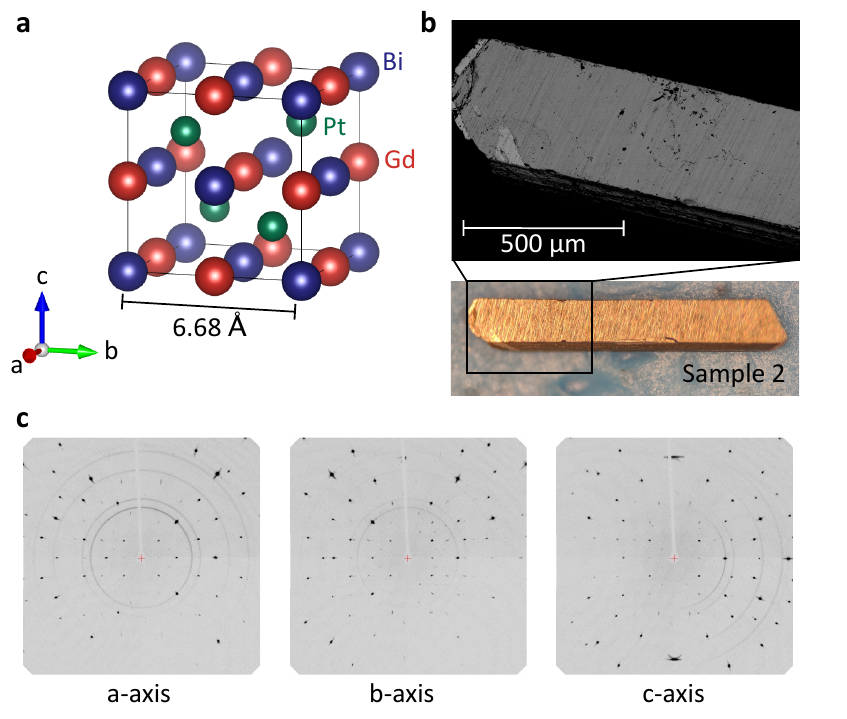}
	\caption{a) Crystallographic unit cell of GdPtBi. b) Bottom: Photograph of single-crystalline GdPtBi (Sample 2) cut to a bar-shaped rod. Top: SEM image. The bright areas in the BSE-mode show Bi-flux inclusions at the edge of the rod, confirmed by EDXS analysis. c) Single crystal X-ray diffraction images for the main crystal directions.}
	\label{fig:crystal}
\end{figure}
GdPtBi crystallizes in a fcc-lattice as shown in Fig.~\ref{fig:crystal}a with a lattice constant of $a=\SI{6.68}{\angstrom}$.
The single-crystalline samples used in our study were grown in Bi self-flux\cite{CanfieldGrowth}.
Two $\sim $ 3\,mm-sized samples could be extracted from two different batches (Sample 1 and Sample 2), which have been cut along crystalline axes with a wire-saw prior to orientation using X-ray diffraction.
The X-ray diffraction pattern (Fig.~\ref{fig:crystal}c) shows the characteristic fcc-reflections; the rings most likely stem from powder contaminations on the surface due to the polishing process.
The stoichiometry of the samples has been confirmed with energy-dispersive X-ray spectroscopy (EDXS) analysis on multiple points on all surfaces of the samples, however, patches of $<100\,\mathrm{\mu m}$ of enclosed Bi-flux are sometimes revealed upon polishing the sample. 
They appear as shiny silver, distinct from the more bronze GdPtBi-phase, and can also be identified as bright patches under the electron microscope in the back-scattered electron mode (Fig.~\ref{fig:crystal}b).
We cut all visible Bi-flux in the investigated GdPtBi-samples away and broke the samples in half after the measurements to check for flux at the edges, however, a residual risk of having undetected enclosures always remains.
The magnetization $M$ was measured in a Quantum Design Magnetic Property Measurement System (MPMS) magnetometer equipped with a 7\,T magnet.
The transport measurements have been performed in a Quantum Design Physical Property Measurement System (PPMS) with a 9\,T magnet.
Electrical and thermal currents, $\bm{J}$ and $\bm{J}_\mathrm{h}$, were applied along the same axes for each sample, which is along [100] for Sample 1 and along [211] for Sample 2.
Electrical resistivity and Hall effect measurements were performed in a 4-probe configuration using Stanford SR-830 lock-in amplifiers with a reference frequency of 77.77\,Hz and an excitation current of $1\,\mathrm{mA}$.
Thermal resistivity measurements were performed with the steady-state method using the commercial Thermal Transport Option (TTO) sample holder from Quantum Design, whereby a $2\,\mathrm{k\Omega}$ Ruthenium oxide resistor is attached to one end of the sample as a heater, two Cernox thermometers are attached in the middle part $\sim 1\,\mathrm{mm}$ apart, and the lower end of the sample is attached to a Cu-block acting as a heat sink.
To achieve good thermal coupling to the sample for thermal resistance measurements, we have fabricated four thermal contacts using $0.6\times0.25\,\mathrm{mm^2}$ gold-plated Cu-bars for the heater and the heat sink; and 0.1\,mm Pt-wire for the thermometers.
In order to avoid thermal smearing, the heater power was set such that the temperature gradient $\Delta T$ does not exceed 3\% of the base temperature $T$.
In addition, in both electrical and thermal measurements homogeneity of the current flow was ensured by covering the facets connected to the current leads with silver paint, as well as attaching the voltage and thermometer leads across the whole sample width at an expense of a geometrical error due to the contact size (not more than 10\%).
The magnetoelectrical resistivity (MR) $\rho_{xx}$ and magneto-thermal resistivity (MTR) $w_{xx}$ have been symmetrized with respect to $\bm{H}=0$, and the Hall resistivity $\rho_{xy}$ antisymmetrized, respectively, to eliminate small misalignment effects.
\section{Sample Characterization}
\label{sec:Characterization}
\begin{figure}[ht!]
	\centering
	\includegraphics[width=8.6cm]{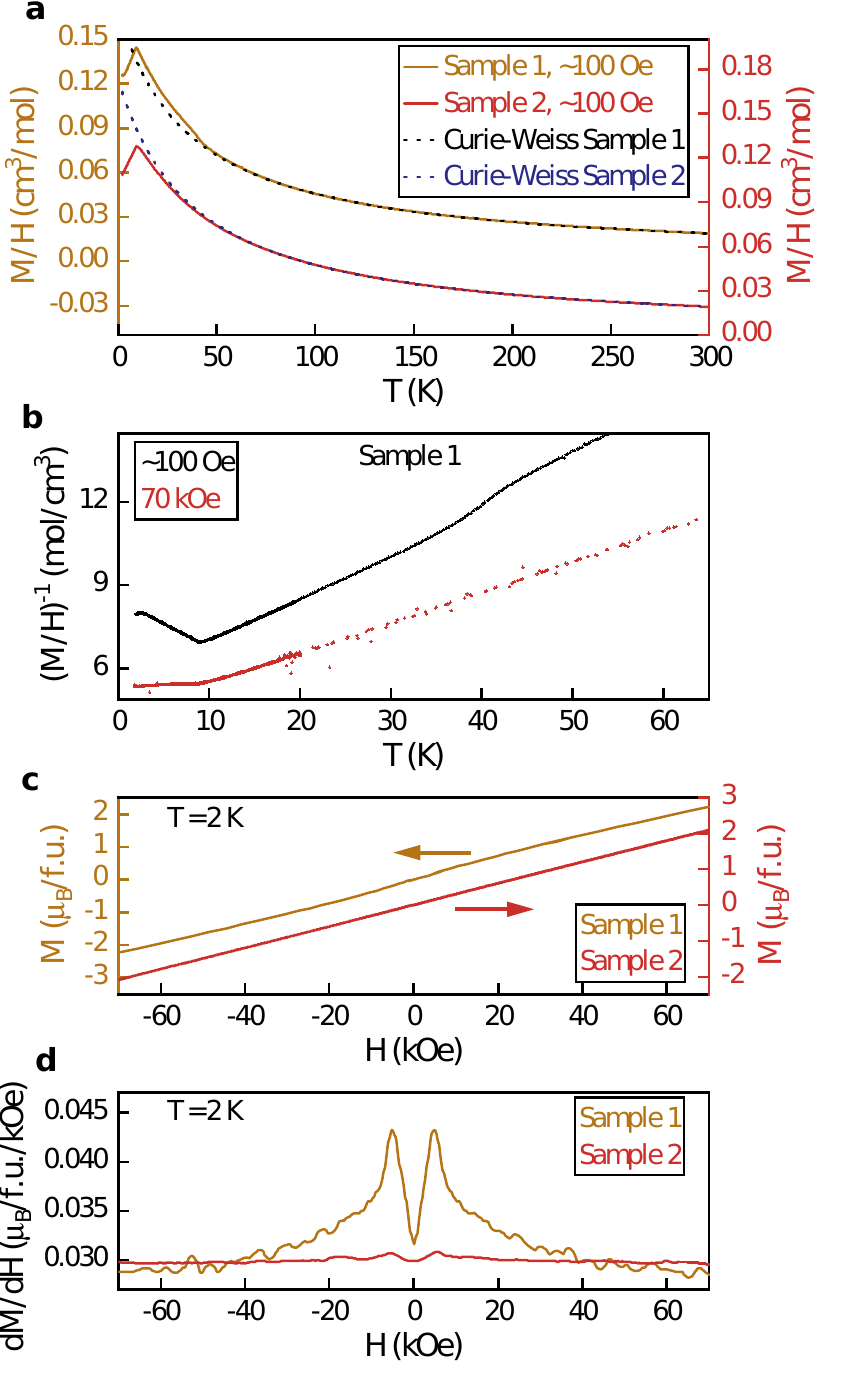}
	\caption{a) Molar susceptibility $M/H(T)$. The dotted lines show the Curie-Weiss law for orientation. b) Inverse susceptibility $(M/H)^{-1}(T)$ of Sample 1 at 100\,Oe and 70\,kOe. c) Magnetization $M(H)$. d) Differential susceptibility $\chi(H)=\mathrm{d}M/\mathrm{d}H(H)$. Sample 1 shows a pronounced non-linearity of $M(H)$ in low fields.}
	\label{fig:Magnetization}
\end{figure}
The magnetic susceptibility $M/H$ at 100\,Oe as a function of temperature $T$ is shown in Fig.~\ref{fig:Magnetization}a.
The antiferromagnetic phase transition is clearly observed for both samples at $T_\mathrm{N}\sim 8.9\,\mathrm{K}$.
Above $T_\mathrm{N}$, Sample 2 follows the Curie-Weiss law\cite{Coey}
\begin{equation}
M/H=\frac{C}{T-\varTheta_\mathrm{CW}}+\chi_0,
\label{eq:CurieWeiss}
\end{equation}
where $C$ is the Curie constant, $\varTheta_\mathrm{CW}$ the Curie temperature and $\chi_0$ is the $T$-independent part of the susceptibility (e.g. Van-Vleck susceptibility).
In contrast to Sample 2, Sample 1 exhibits a positive deviation from the Curie-Weiss law\cite{Coey} below 44\,K, indicating the presence of a minor ferromagnetic impurity phase.
The extra signal is not visible in a higher applied field, most apparent in the inverse susceptibility curve (see Fig.~\ref{fig:Magnetization}b), supporting the ferromagnetic character of the impurity.
The transition is not very sharp, thus it most likely stems from a Gd-Pt alloy phase, since they all order ferromagnetically near 44\,K (e.g. GdPt at 68\,K\cite{magnetostriction}, $\mathrm{GdPt_{2-x}}$ at 36-46.5\,K\cite{Buschow1979}, $\mathrm{GdPt_5}$ at 13.9\,K\cite{Smith1981}).
We estimate that 30\,ppm of ferromagnetic impurity phase is sufficient to explain the strength of the observed signal, thus detection via EDXS is below the achievable resolution.
We note that there is a systematic field error of the setup of up to 20\,Oe, causing significant deviation of the absolute susceptibility values at low fields.
We therefore fitted the $M/H(T)$ curve recorded at 70\,kOe in the range 30-300\,K for extracting of the Curie-Weiss parameters.
With $C=1.571\times10^{-6}g^2J(J+1)$ \cite{Coey}, the effective paramagnetic moment $m_\mathrm{eff}=g\mu_\mathrm{B}\sqrt{J(J+1)}$ can be estimated, yielding $(8.00\pm0.10)\mu_\mathrm{B}$ for Sample 1 and $(8.08\pm0.04)\mu_\mathrm{B}$ for Sample 2.
These values are slightly larger than the moment of the free $\mathrm{Gd^{3+}}$ ion $7.94\mu_\mathrm{B}$ (with $g=2$ and $J=7/2$).
The extracted Curie temperatures are $(-38.2\pm0.5)\mathrm{K}$ for Sample 1 and $(-40.2\pm0.2)\mathrm{K}$ for Sample 2, yielding a moderate frustration parameter\cite{Ramirez1994} $f=-\varTheta_\mathrm{CW}/T_\mathrm{N}\sim 4$.
The Curie-Weiss parameters of our samples are in good agreement with previous reports\cite{Suzuki2016,Canfield1991,Mueller2014,Hirschberger2017,Shekhar2016,Mun2016},
The slight upturn of $M/H(T)$ below 2.5\,K indicates the presence of a paramagnetic impurity in Sample 1, additional to the ferromagnetic impurity.
It is also manifested in the $M(H)$ curve (Fig.~\ref{fig:Magnetization}b), where Sample 1 shows a field-symmetric non-linearity around 1\,T, better visible in the derivative $\chi(H)=\mathrm{d}M/\mathrm{d}H(H)$ (Fig.~\ref{fig:Magnetization}c).
Again, we could not clarify the origin of the paramagnetic impurity in Sample 1, as it falls below the resolution of our EDXS.
Ultimately, enclosed Bi-flux can be excluded as an origin, since it would just enhance the diamagnetic contribution.
We note that similar signatures in the $M(H)$-curves have been reported for GdPtBi samples in ref.~\onlinecite{Hirschberger2017}, suggesting that paramagnetic impurities are a common issue when growing GdPtBi single-crystals. 
\begin{figure}[ht!]
	\centering
	\includegraphics[width=8.6cm]{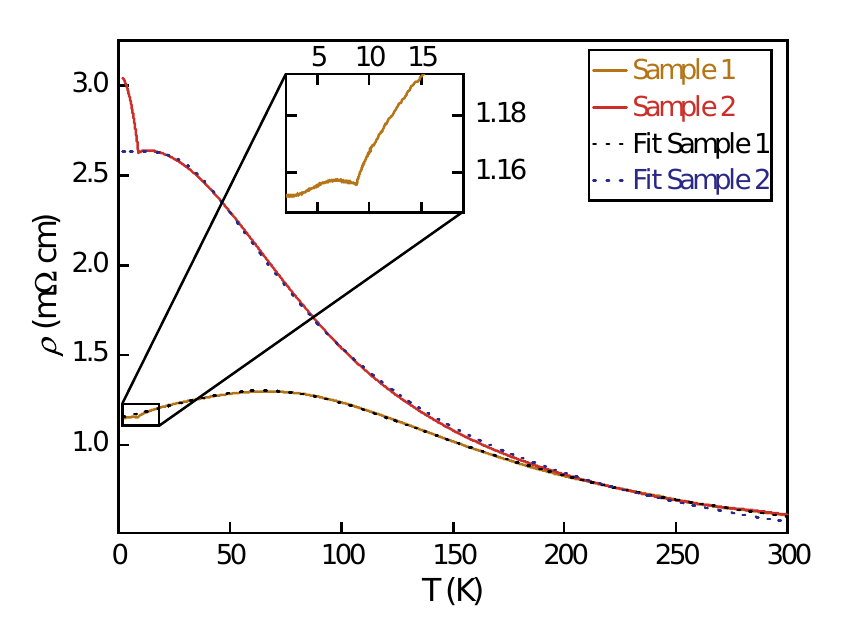}
	\caption{Zero-field resistivity $\rho(T)$. For Sample 1, the region around the antiferromagnetic transition at $T_\mathrm{N}\sim 9\,\mathrm{K}$ is enlarged in the inset. The dashed lines show fits with Eq.~\ref{eq:tempcurve}.}
	\label{fig:RvsT}
\end{figure}
\begin{figure*}[ht!]
	\centering
	\includegraphics[width=17.2cm]{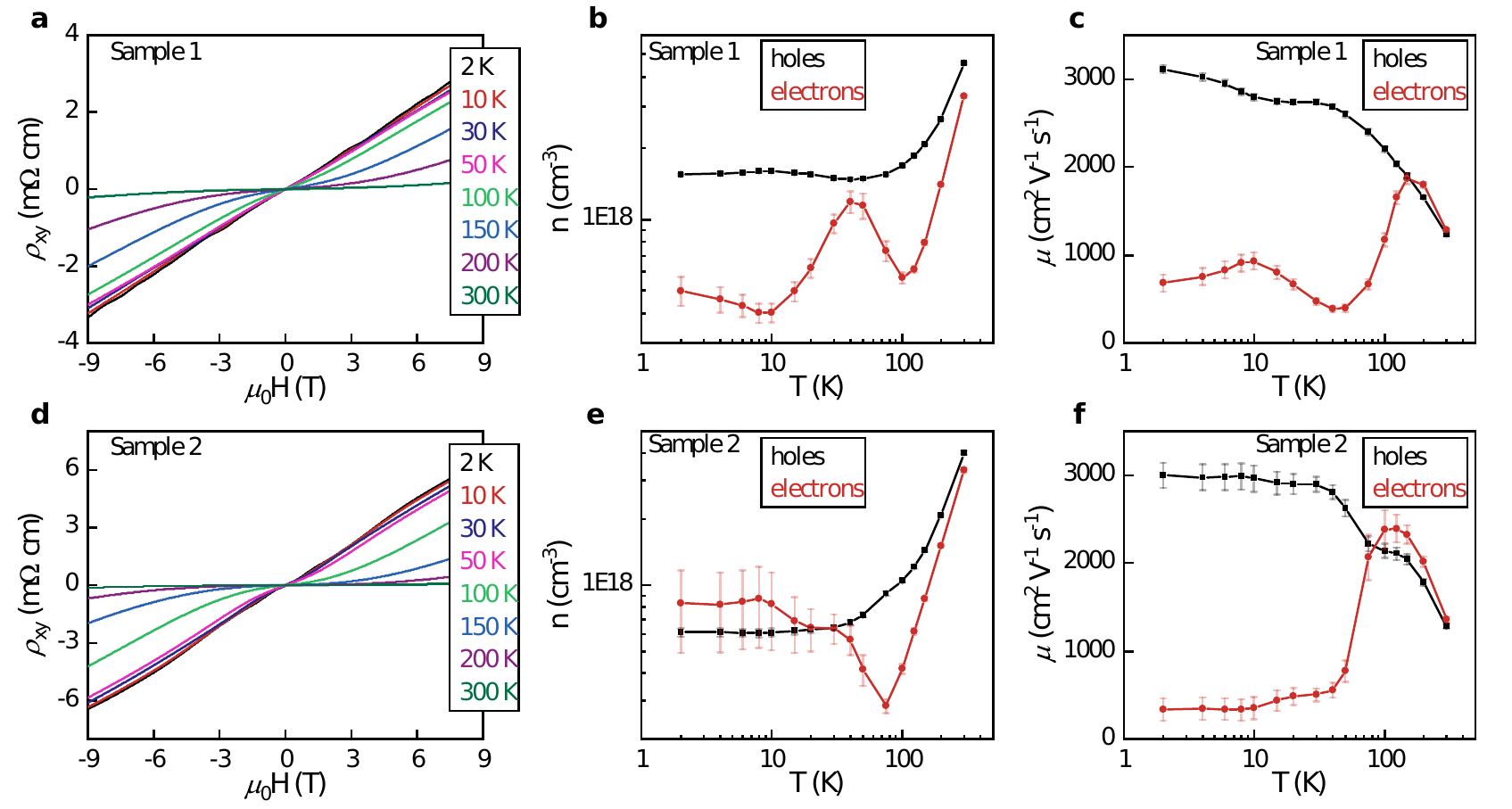}
	\caption{(a,d) Hall resistivity $\rho_{xy}(H)$ for $\bm{H}\perp\bm{J}$ for Samples 1 (a) and 2 (d). (b,e) Carrier densities $n$ of holes and electrons extracted by fitting $\rho_{xy}(H)$ with Eq.~\ref{eq:twocarrierHall}. (c,f) Extracted Drude mobilities $\mu$ from the fit with Eq.~\ref{eq:twocarrierHall}.}
	\label{fig:Carrier}
\end{figure*}
The electrical resistivity $\rho(T)$ of both samples shows the behavior expected for a zero-gap semimetal (Fig.~\ref{fig:RvsT}).
The sharp cusp at $T_\mathrm{N}\approx 9\,\mathrm{K}$ has been reported to occur due to the formation of a small superzone gap of 1\,meV at the antiferromagnetic ordering of Gd's $4f$-moments\cite{Mueller2014,Suzuki2016,Kreyssig2011,Canfield1991,Mun2016,Hirschberger2017}.
As stated in ref.~\onlinecite{Hirschberger2016}, the robustness of $T_\mathrm{N}$ against different doping levels suggests that the conduction carriers are not involved in the coupling mechanism of the Gd-moments.
However, the reverse implication, that the conduction carriers are not affected by the localized Gd-moments, is not necessarily the case as the cusp in $\rho(T)$ signifies.
Sample 2 exhibits a strong increase in $\rho(T)$ below $T_\mathrm{N}$, indicating a Fermi level situated closer to the band touching point compared to Sample 1.
Above $T_\mathrm{N}$, $\rho(T)$ can be fitted with a simplified two-carrier model, where the Fermi level is positioned at the edge of a hole band and electrons are thermally excited across $E_\mathrm{g}$, with $E_\mathrm{g}$ being the energy difference of the electron band minimum and the Fermi level.
In that way, $E_\mathrm{g}$ can be understood as a measure for the energy distance of the Fermi level from the band touching point.
The total number of carriers is modeled as\cite{Berger2002}
\begin{equation}
n(T)=n_0+N\sqrt{k_\mathrm{B}T\ln{2}[k_\mathrm{B}T\ln{(1+\mathrm{e}^{E_\mathrm{g}/k_\mathrm{B}T})}-E_\mathrm{g}]},
\end{equation}
where $n_0$ is the number of temperature-independent carriers and $N$ the density of states of the individual bands.
The resistivity then holds\cite{Berger2002}
\begin{equation}
\rho(T)=\frac{\rho_0 n_0+A T}{n(T)},
\label{eq:tempcurve}
\end{equation}
where the term $A T$ accounts for the phonon scattering contribution, which is assumed to be linear above $\Theta_\mathrm{D}/10$, where $\Theta_\mathrm{D}$ is the Debye temperature.
The fits (see Fig.~\ref{fig:RvsT}) yield $E_g=(42\pm3)\mathrm{meV}$ for Sample 1 and $(15\pm5)\mathrm{meV}$ for Sample 2, which gives a rough estimate of how close the Fermi level is positioned to the band touching point and thus the size of the Fermi sea.
Consistently, the Hall resistivity $\rho_{xy}(H)$ exhibits an increasingly non-linear slope upon increasing $T$ (see Fig. \ref{fig:Carrier}a,d).
To extract carrier mobilities $\mu$ and densities $n$ of the individual hole and electron bands, the Hall data was fitted with a two-carrier Drude model\cite{Honig1963}
\begin{equation}
\rho_{xy}=\frac{1}{D}\Big[R_1\sigma_1^2+R_2\sigma_2^2+R_1R_2\sigma_1^2 \sigma_2^2(R_1+R_2)H^2\Big],
\label{eq:twocarrierHall}
\end{equation}
whilst fitting the transverse MR using the same parameters with
\begin{equation}
\rho_{xx}=\frac{1}{D}\Big[(\sigma_1+\sigma_2)^2+\sigma_1 \sigma_2(\sigma_1 R_1^2+\sigma_2 R_2^2)H^2\Big],
\label{eq:two-carrierMR}
\end{equation}
with $D=(\sigma_1+\sigma_2)^2+(\sigma_1 \sigma_2)^2(R_1+R_2)^2 H^2$, where $\sigma_{1,2}=(nq\mu)_{1,2}$ are the Drude conductivities and $R_{1,2}=(1/qn)_{1,2}$ the Hall coefficients of the individual electron/hole channels.
The extracted carrier densities are displayed in Figs.~\ref{fig:Carrier}b,e; the mobilities in Figs.~\ref{fig:Carrier}c,f.
In both samples, hole-type transport is dominant, yielding carrier densities in the order of $10^{18}\,\mathrm{cm^{-3}}$ for Sample 1 and $10^{17}\,\mathrm{cm^{-3}}$ for Sample 2, consistent with a Fermi level closer to the band touching point and thus a smaller Fermi sea in Sample 2.
This equates to only $\sim 10^{-4}$ majority carriers per unit cell.
Mobilities and carrier densities of the electron pockets are much less reliably determined, however, either density or mobility of the electrons are much lower than the hole values; or both.
\begin{figure}[ht!]
	\centering
	\includegraphics[width=8.6cm]{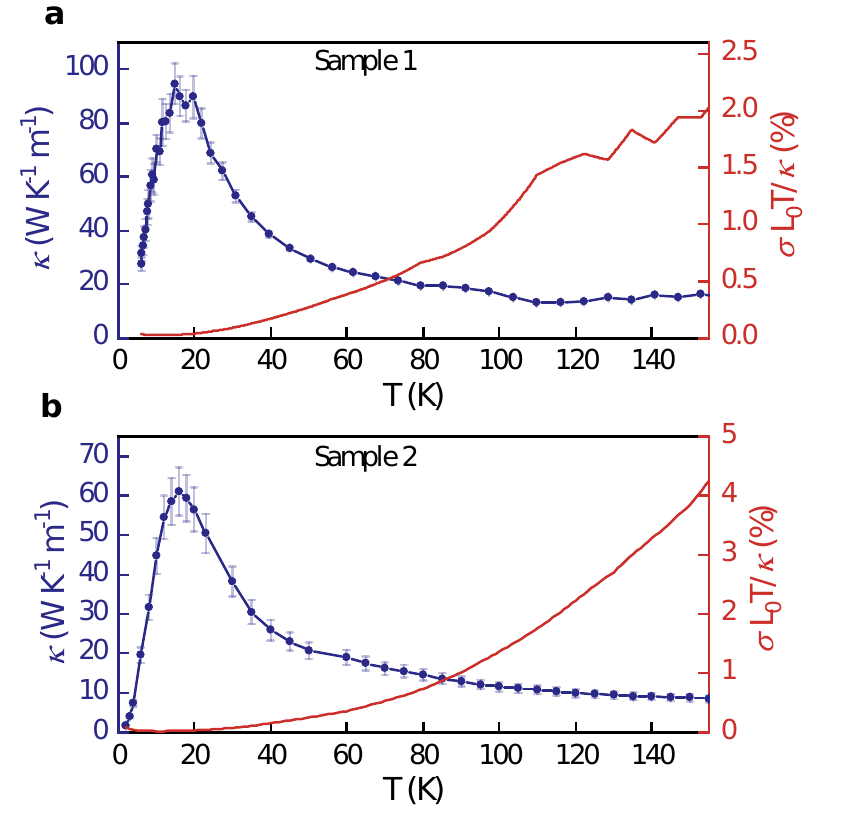}
	\caption{Thermal conductivity $\kappa(T)$ of Sample 1 (a) and Sample 2 (b). The left axes show $\kappa(T)$, the right axes show the estimated electronic contribution $\kappa_\mathrm{el}$ from the Wiedemann-Franz law.}
	\label{fig:Thermal}
\end{figure}
The zero-field temperature dependence of the thermal conductivity $\kappa(T)$ (left axes in Fig.~\ref{fig:Thermal}a,b)  is comparable to the $\kappa(T)$-curves reported in refs.~\onlinecite{Hirschberger2016,Hirschberger2017}.
Concerning the low density of electronic carriers, it is not surprising that the overall $T$-dependence of the thermal transport in GdPtBi can be fully explained by phonon conduction: upon warming from 2\,K, $\kappa(T)$ increases with $T^3$ due to the increasing lattice heat capacity.
Near 15\,K, $\kappa(T)$ reaches a maximum of around $95\,\mathrm{WK^{-1}m^{-1}}$ in Sample 1 ($60\,\mathrm{WK^{-1}m^{-1}}$ in Sample 2) and then starts to decrease exponentially due to the onset of phonon Umklapp scattering.
At higher $T$, the exponential drop is replaced by a slower $1/T$ power law-dependence.
Estimating the electronic contribution of $\kappa$ via the Wiedemann-Franz law $\kappa_\mathrm{el}\approx L_0 T/\rho$, with the Lorenz number $L_0=2.44\cdot10^{-8}\,\mathrm{W\Omega K^{-2}}$, it is apparent that thermal transport in GdPtBi is dominated by the lattice contribution.
The ratio $\kappa_\mathrm{el}/\kappa$ is illustrated on the right axis in Fig.~\ref{fig:Thermal}.
At 150\,K, the estimated electronic contribution still contributes $\sim$2-5\%, however, it decreases strongly to less than 0.1\% below 30\,K.
We note that violations of the Wiedemann-Franz law due to strong correlations as well as carrier compensation usually result in a reduced Lorenz number\cite{Pfau2016,Li2018,Gooth2018}, and we are not aware of any case (except for quasi-one-dimensional conductors) where Lorenz numbers larger than $25L_0$ are observed\cite{Crossno2016}.
Even if such gross violation of the Wiedemann-Franz law would be the case, we can still assume that the electronic contribution at low temperatures is negligible.
\section{Band structure and Shubnikov-de Haas oscillations}
\label{sec:Bandstructure}
\begin{figure}[ht!]
	\centering
	\includegraphics[width=8.6cm]{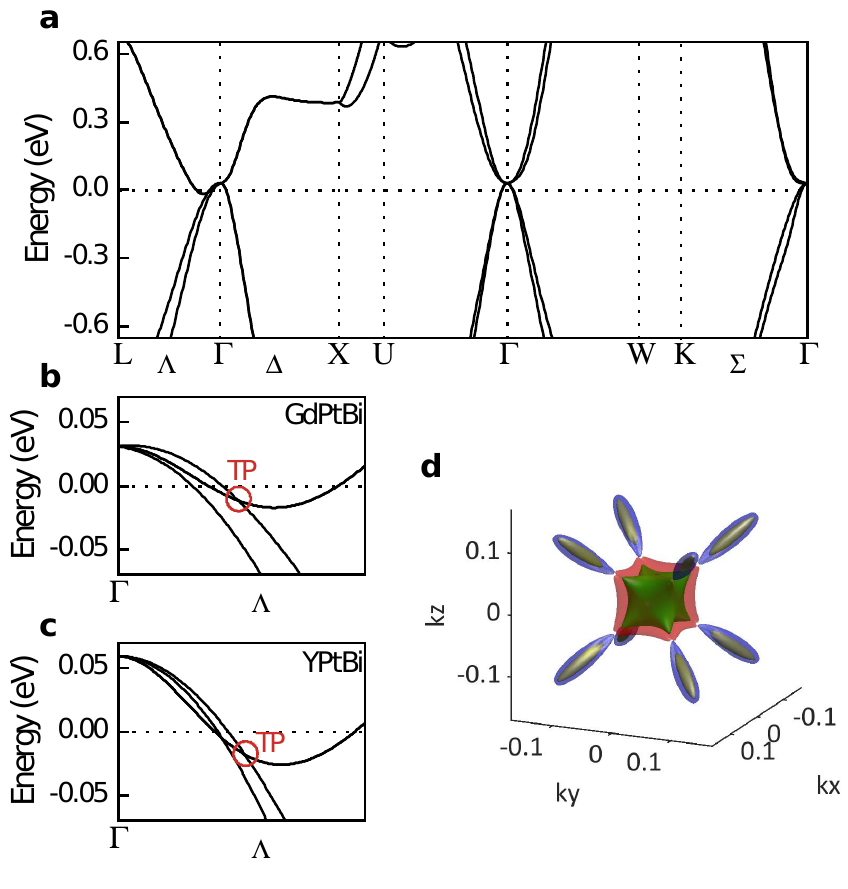}
	\caption{a) Electronic band structure of GdPtBi calculated from DFT. b) and c) Triple points along the $\Gamma-\mathrm{L}$-line in GdPtBi and YPtBi. d) Fermi surface of GdPtBi. The intercalated red and green cube-shaped pockets are hole-like, the blue and yellow tear-shaped pockets are electron-like.}
	\label{fig:Bandstructure}
\end{figure}
We performed DFT based first principles calculations using the code of Vienna \emph{ab-initio} simulation package (VASP) with the projected augmented wave method~\cite{vasp1,vasp2}. 
The exchange and correlation energies were considered in the generalized gradient approximation (GGA) with the Perdew-Burke-Ernzerhof-based density functional~\cite{pbe}; the $4f$-electrons in Gd were considered as the core state.
We have projected the Bloch wavefunctions into the maximally localized Wannier functions (MLWFs)~\cite{wannier} and constructed the effective tight binding model Hamiltonian.
The resulting band structure (Fig.~\ref{fig:Bandstructure}a) is in good agreement with previous reports\cite{Suzuki2016,Chadov2010} and conforms to the picture of a zero-gap semimetal with touching conduction and valence bands at the $\Gamma$-point.
Consistent with previous calculations\cite{Barik2018} and recent optical measurements on GdPtBi\cite{Hutt2018}, our model yields a triply degenerated point near the $\Gamma$-point along the $\Gamma$-L line (Fig.~\ref{fig:Bandstructure}b), situated around 10\,meV below the Fermi level.
To check, whether the triple point crossings are exclusive to GdPtBi and can thus be considered as an origin for the anomalous magnetotransport features\cite{Lepori2018}, we also calculated the band structure for the non-magnetic YPtBi (Fig.~\ref{fig:Bandstructure}c).
A similar triple point near the Fermi level is found, yet no signatures of anomalous magnetotransport have been reported for YPtBi \cite{Shekhar2016}, thus indicating the insignificance of the triple point crossings for electronic transport in GdPtBi .
The calculated Fermi surface (Fig.~\ref{fig:Bandstructure}d) shows two intercalated cube-shaped hole pockets centered around the $\Gamma$-point, and 8 pairs of small, teardrop-shaped electron pockets at the corners of the hole pockets.
We note that the size and shape of the pockets sensitively depends on the position of the Fermi level; when it is sufficiently low, the electron pockets fully disappear.
\begin{figure*}[ht!]
	\centering
	\includegraphics[width=17.2cm]{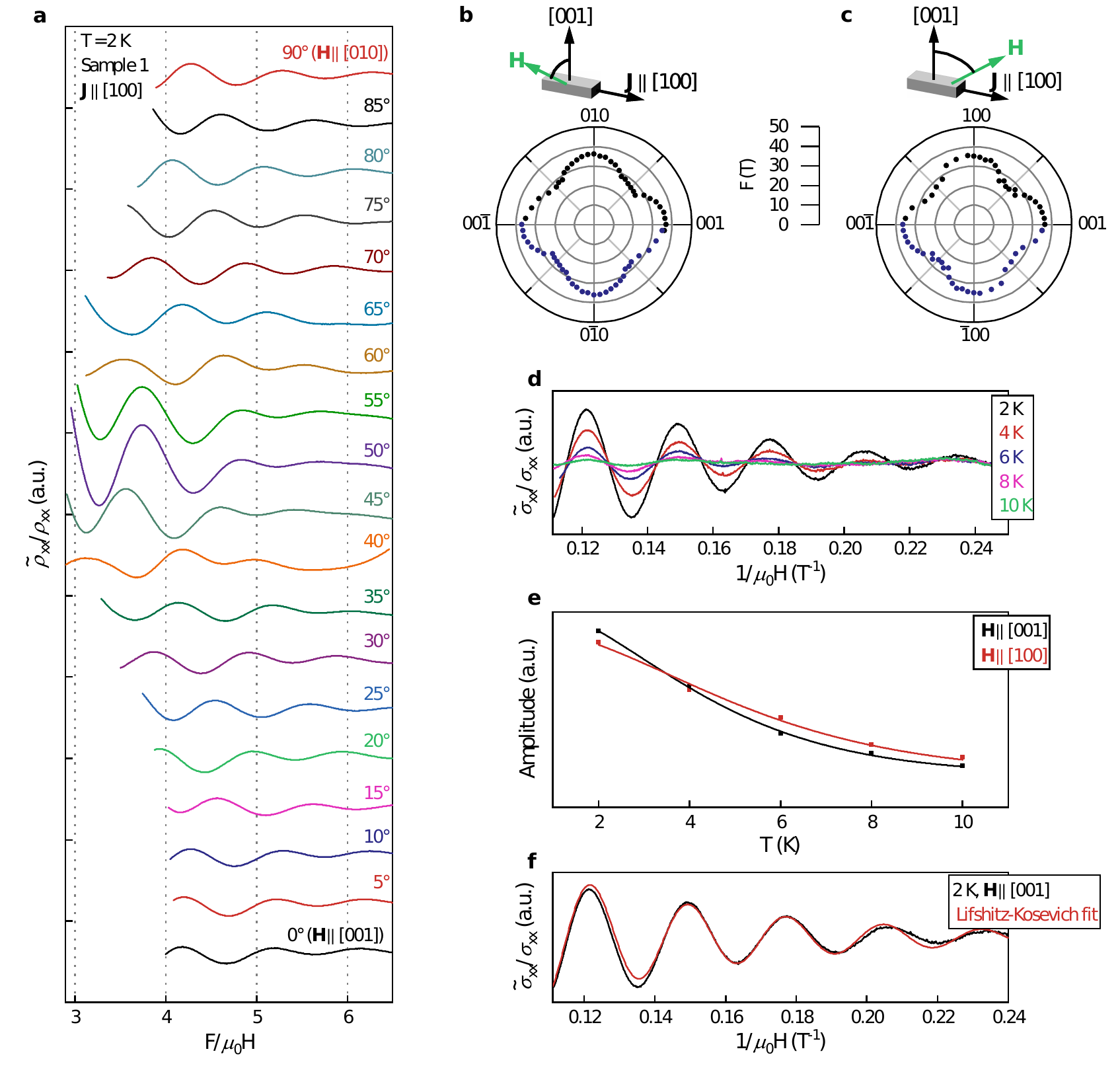}
	\caption{Shubnikov-de Haas (SdH) oscillations in Sample 1. a) Extracted oscillatory part $\tilde{\rho}_{xx}/\rho_{xx}$ versus $F/H$ for applied magnetic fields $\bm{H}$ in the [100]-plane, rotating from [001] to [010]. (b,c) SdH frequency $F$ versus direction of $\bm{H}$ for rotation in the [100]-plane (b) and the [010]-plane (c). The black dots are extracted from the experimental data, the blue dots are duplicated according to symmetry. d) $\tilde{\sigma}_{xx}/\sigma_{xx}$ versus $1/H$ for different temperatures $T$. e) Temperature-damping of $\tilde{\sigma}_{xx}/\sigma_{xx}$ at fixed $1/H$. The solid lines represent fits with Eq.~\ref{eq:tempdamping}. f) Fit of the SdH oscillation at 2\,K with the Eq.~\ref{eq:LK}.}
	\label{fig:SdH}
\end{figure*}
Sample 1 showed clear Shubnikov-de Haas (SdH) oscillations in the MR at moderate magnetic fields, allowing for an experimental reconstruction of the Fermi surface.
The oscillations in $\rho_{xx}$ are periodic in $1/H$ with a frequency\cite{Shoenberg}
\begin{equation}
	F = \frac{\hbar}{2\pi e} A_\mathrm{ext},
	\label{eq:Onsager}
\end{equation}
$A_\mathrm{ext}$ being the extremal cross-sectional area of the Fermi surface in the plane perpendicular to the applied magnetic field $\bm{H}$.
The oscillatory part $\tilde{\rho}_{xx}/\rho_{xx}$ was extracted by subtracting a smooth polynomial background from the $\rho_{{xx}}$-curve and plotting the residual versus $1/H$.
By assigning resistivity maxima (and minima) with (half-)integers and fitting a linear function, $F$ can be extracted from the slope.
The extracted SdH oscillations versus $F/H$ are displayed for several fields rotated in the plane perpendicular to $\bm{J}\parallel [100]$ in Fig.~\ref{fig:SdH}a.
Figs.~\ref{fig:SdH}b,c show the variation of $F$ upon rotation of $H$ in the plane perpendicular to [100] and [010], respectively.
The frequencies have been extracted within a $180^\circ$ range, as the remaining range should be symmetric (duplicated values displayed in blue in Figs.~\ref{fig:SdH}b,c).
The four-fold rotational symmetry of the underlying crystal lattice becomes apparent, as $F$ undergoes a maximum of $\sim$35\,T when $\bm{H}$ is aligned along one of the main axes, and a minimum of $\sim$26\,T at $45^\circ$ from the main axes.
As expected, the rotations in the [100] and [010] planes yield nearly identical frequencies.
Assuming a cube-shaped pocket, the volume enclosed by the Fermi surface can be estimated via $A_\mathrm{ext}^{3/2}$, $A_\mathrm{ext}=(3.3\pm0.2)\,10^{-3}\si{\angstrom^{-2}}$ extracted via Eq.~\ref{eq:Onsager} for $\bm{H}\parallel [100]$.
The carrier density can then be estimated with $n=2/(2 \pi )^3\,A_\mathrm{ext}^{3/2}=(1.53\pm0.10)\,10^{18}\,\mathrm{cm^{-3}}$.
This is in agreement with the hole density extracted from the Hall measurement $(1.54\pm0.02)\,10^{18}\,\mathrm{cm^{-3}}$ at 2\,K.
Conclusively, the SdH oscillations stem from the hole pocket.
In contrast to the DFT results (see Fig.~\ref{fig:Bandstructure}d), where at least two distinct SdH frequencies stemming from the two differently sized hole pockets would be expected, only one frequency is observed.
This indicates that the single-electron picture used for calculating the band structure does not fully account for the actual shape of the Fermi surface and many-body effects might need to be included in the theoretical treatment for better agreement between theory and experiment\cite{Chadov2010}.
For further analysis, the magnetoconductivity $\sigma_{xx}=\rho_{xx}/(\rho_{xx}^2+\rho_{xy}^2)$ has been calculated, as the oscillatory part $\tilde{\sigma}_{xx}/\sigma_{xx}$ is proportional to the oscillations in the density of states and can be described by the Lifshitz-Kosevich formalism\cite{Shoenberg}.
In the simplest form (neglecting many-body interactions) it holds\cite{Shoenberg}
\begin{equation}
	\frac{\tilde{\sigma}_{xx}}{\sigma_{xx}}\propto \sum_{p=1}^{\infty} \left(\frac{H}{p}\right)^{1/2} R_\mathrm{D} R_\mathrm{T} R_\mathrm{s} \cos\left[2\pi p \left(\frac{F}{H}-\beta\right)\pm\frac{\pi}{4}\right],
	\label{eq:LK}
\end{equation}
where the phase factor $\beta$ depends on the details of the band structure and the $\pm$ accounts for the extremal area being a mimimum (+) or a maximum (-).
The Dingle damping factor $R_\mathrm{D}$ accounts for a finite relaxation time and holds
\begin{equation}
	R_\mathrm{D}=\exp\left[-p\frac{\pi}{\mu_\mathrm{c} H}\right],
	\label{eq:Dingle}
\end{equation}
where $\mu_\mathrm{c}$ is the mobility in case of a semiclassical cyclotron motion.
The damping factor $R_\mathrm{T}$ accounts for a finite temperature and holds
\begin{equation}
R_\mathrm{T}=\frac{\lambda (T)}{\sinh[\lambda(T)]},\mathrm{with}~\lambda(T)=2\pi^2 p\frac{m_\mathrm{c}}{ eH}\frac{k_\mathrm{B}T}{\hbar},
\label{eq:tempdamping}
\end{equation}
where $m_\mathrm{c}$ is the cyclotron mass.
The damping factor $R_\mathrm{s}$ accounts for the effect of spin-splitting and holds
\begin{equation}
	R_\mathrm{s}=\cos\left(\frac{1}{2}p\pi g \frac{m_\mathrm{c}}{m_0}\right),
\end{equation}
where $g$ is the Land\'{e}-factor and $m_0$ the free electron mass.
The extracted oscillatory part of $\sigma_{xx}$ is shown in Fig.~\ref{fig:SdH}d for different temperatures.
The characteristic damping of the amplitude allows to extract $m_\mathrm{c}$ by fitting $\tilde{\sigma}_{xx}/\sigma_{xx}$ versus $T$ at fixed $1/H$ with Eq.~\ref{eq:tempdamping} (see Fig.~\ref{fig:SdH}e).
It has been determined to $m_\mathrm{c}=(0.30\pm0.02)m_0$ for $\bm{H}\parallel[100]$, and $m_\mathrm{c}=(0.26\pm0.02)m_0$ for $\bm{H}\parallel[001]$.
These results are similar to those reported in refs.~\onlinecite{Hirschberger2016,Hirschberger2017}.
A fit of the full SdH oscillation at 2\,K with Eq.~\ref{eq:LK} is shown in Fig.~\ref{fig:SdH}f, only taking the fundamental oscillation ($p=1$) into account.
The extracted mobility $\mu_\mathrm{c}=(3200\pm100)\,\mathrm{cm^2 V^{-1} s^{-1}}$ is comparable to the Drude mobility extracted from the Hall analysis $\mu=(3000\pm200)\,\mathrm{cm^2 V^{-1} s^{-1}}$.
The extraction of the $g$-factor would be very useful for the investigation of the magnetic interactions in GdPtBi.
However, we were not able to draw any conclusions about the spin-splitting\cite{Shoenberg}, since the higher harmonics ($p>1$) are dampened too strongly in the field range accessible in our setup.
High-field MR measurements on GdPtBi have been performed by Hirschberger\cite{Hirschberger2017}, showing a frequency change above a critical field of $\mu_0 H_\mathrm{c}\sim 25\,\mathrm{T}$ that could not be associated with a spin-splitting.
At this critical field, also the $M(H)$-curves below $T_\mathrm{N}$ are reported\cite{Hirschberger2017,Suzuki2016} to show a clear kink and slope change, attributed to the saturation of the magnetic moments which potentially leads to strong alterations of the band structure\cite{Hirschberger2017}.
Indications for field-induced modifications of the band structure have also been observed in the antiferromagnetic sister compound CePtBi\cite{Kozlova2005}, where the amplitude of the SdH oscillations drastically changes above the critical field of 25\,T. 
In Fig.~\ref{fig:SdH}a it becomes apparent that the phase of the SdH oscillations in GdPtBi is strongly varying upon rotation of $\bm{H}$, which has also been reported in ref.~\onlinecite{Hirschberger2017}.
Possible mechanisms for this phase variation can be\cite{Shoenberg}:
(i) the variation of the extremal orbit from maximum to minimum orbits causing a phase jump of $\pm \pi/4$ according to Eq.~\ref{eq:LK},
(ii) the peculiarities of the band structure causing a deviation of $\beta$ from the value $1/2$ for parabolic bands,
(iii) the strength of the spin-splitting and interactions with the magnetic ions.
\section{Electrical Magnetotransport}
\label{sec:MR}
\begin{figure*}[ht!]
	\centering
	\includegraphics[width=17.2cm]{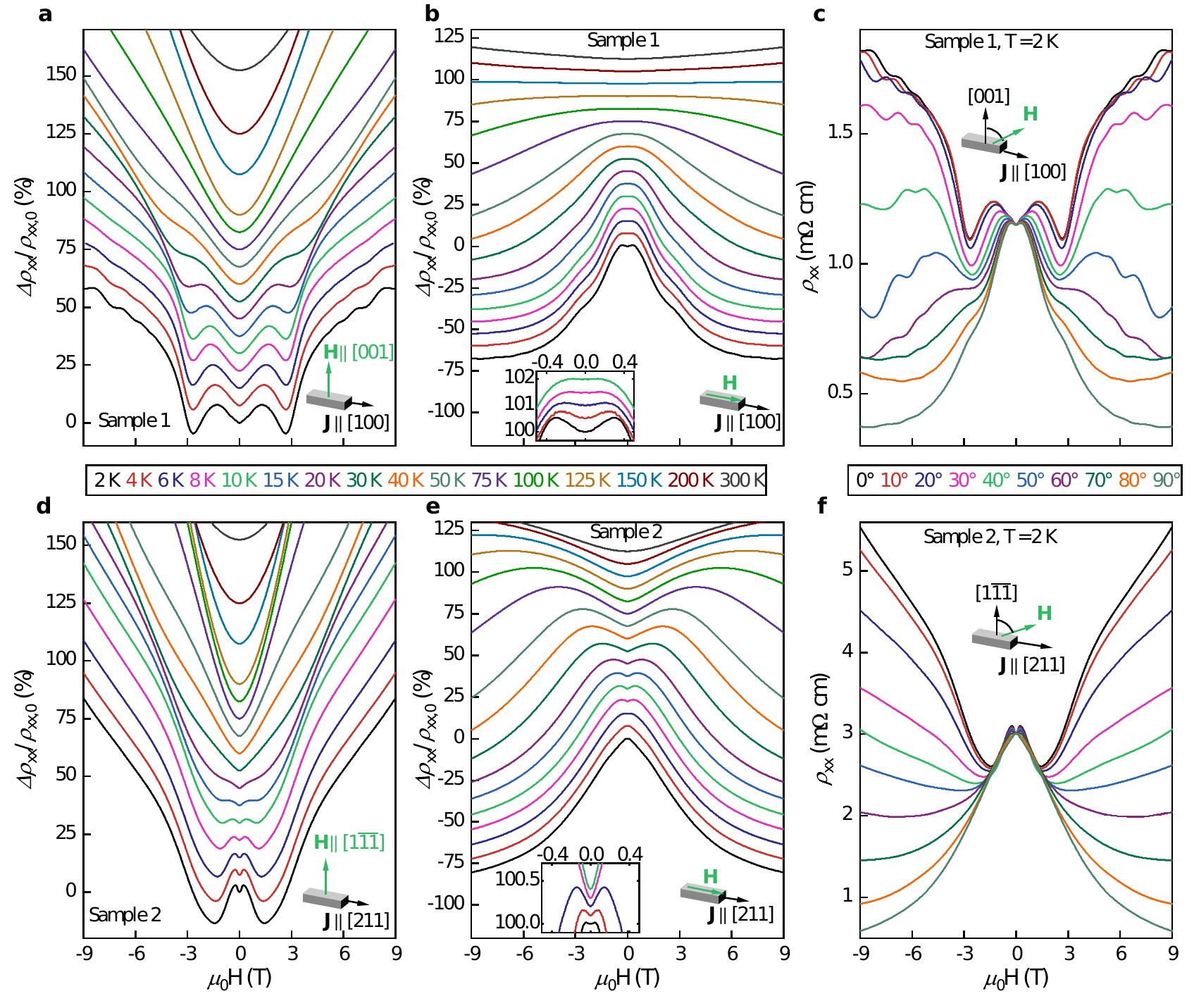}
	\caption{Anisotropic magnetoresistance (MR) $\Delta\rho_{xx}(H)/\rho_{xx}(0)$ for Samples 1 and 2. (a,d) Transverse MR ($\bm{H}\perp\bm{J}$) for different temperatures. (b,e) Longitudinal MR (LMR) ($\bm{H}\parallel\bm{J}$). The insets enlarge the low-field range of the LMR-curves below 10\,K. The curves in (a,b,d,e) are shifted for better visibility. (c,f) MR $\rho_{xx}(H)$ upon rotation of $\bm{H}$ from perpendicular to longitudinal configuration at 2\,K.}
	\label{fig:MR}
\end{figure*}
The transverse MR (TMR) ($\bm{H}\perp\bm{J}$) for different temperatures is displayed in Fig.~\ref{fig:MR}a for Sample 1 and Fig.~\ref{fig:MR}d for Sample 2.
Above 100\,K, the TMR can be adequately described by the two-carrier model (Eq.~\ref{eq:two-carrierMR}), whereby compensation of electron and holes yields a positive TMR with quadratic $H$-dependence in low fields.
Below 100\,K, a pronounced dip evolves around 1-4\,T which deepens upon cooling down and ultimately leads to a negative TMR over a field range of $\sim$2\,T at 2\,K.
Although the exact shape is different for both samples, the main characteristics of the dip are similar and have also been consistently reported for GdPtBi despite varying doping levels\cite{Hirschberger2016,Suzuki2016,Shekhar2016,Mun2016}.
The longitudinal MR (LMR) ($\bm{H}\parallel\bm{J}$) for Sample 1 and 2 is shown in Figs.~\ref{fig:MR}b,e.
Both samples exhibit a bell-shaped negative LMR at low temperatures, at 9\,T and 2\,K resulting in $20-30\%$ of the zero-field value.
The negative LMR weakens upon warming up and ultimately turns positive (in the accessible field range) around 100-150\,K.
The low-field behavior is different for both samples:
At 2\,K, Sample 1 shows a dip around zero-field, which disappears upon warming and leads to a plateau-shape in the low-field LMR curve above 10\,K;
Sample 2 exhibits a dip around zero-field as well, however, upon warming it becomes more pronounced, resulting in a `M'-shaped LMR-curve up to 100\,K.
The strong sample variation of the LMR-shapes can also be seen by comparing the LMR data of previous reports\cite{Shekhar2016,Hirschberger2016}.
Additionally, the low-$T$ LMR curves of Sample 1 exhibit a noticeable feature around 2-3\,T, in a similar field range where the pronounced feature in the TMR appears.
Such a feature also appears in the LMR of the GdPtBi sample in ref.~\onlinecite{Shekhar2016}.
The evolution of this feature upon rotation of $\bm{H}$ suggests a similar origin as the dip in the TMR (Figs.~\ref{fig:MR}c,f).
The overall negative LMR, which successively disappears upon rotating $\bm{H}$ towards the TMR configuration, has previously been associated with the chiral anomaly\cite{Hirschberger2016,Shekhar2016}.
Initially\cite{Hirschberger2016}, the Weyl node creation was thought to be induced by conventional Zeeman splitting, requiring a large spin-orbit coupling with a $g$-factor of the order of $\sim$40 to explain the persistence of the effects up to 150\,K.
However, the absence of the negative LMR and/or anomalous features in the TMR in the non-magnetic sister compounds LuPtBi\cite{Hou2015,Mun2016}, LaPtBi\cite{Kozlova2005} and YPtBi\cite{Shekhar2016,Kim2018} despite a similar band structure strongly indicated that conventional Zeeman-splitting cannot account for the observed effects and that they are related to the unfilled $f$-shell of the rare-earth element. 
Therefore, magnetic exchange enhancing the spin-splitting has been suggested to be responsible for a potential Weyl node creation\cite{Shekhar2016}.
The energy scale of the exchange interaction has been estimated by $g J \mu_\mathrm{B} H_\mathrm{c}$ with $g=2$ and $J=7/2$, yielding $\sim$120\,K\cite{Suzuki2016}.
This is congruent with a magnetic origin causing the appearance of anomalous MR features below this temperature.
\begin{figure}[ht!]
	\centering
	\includegraphics[width=8.6cm]{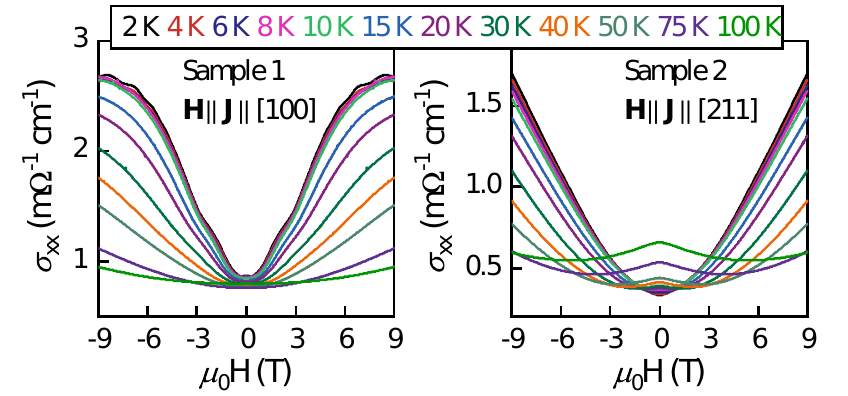}
	\caption{Longitudinal magnetoconductivity $\sigma_{xx,\parallel}=1/\rho_{xx,\parallel}$ for different temperatures.}
	\label{fig:conductivity}
\end{figure}
The transport coefficients in a chiral anomaly scenario\cite{Son2013} yield a positive quadratic function in $H_\parallel$ for $\sigma_{xx}$ in the limit of $H\rightarrow 0$ and a positive linear function in $H_\parallel$ for $H\rightarrow \infty$.
The contribution is tied to the scalar product of $\bm{E}\cdot \bm{H}$, i.e. only $H_\parallel=H\cos(\phi)$ with $\phi=\sphericalangle(\bm{H},\bm{J})$ contributes to the anomalous current generation. 
In agreement with the chiral anomaly scenario, the longitudinal magnetoconductivity $\sigma_{xx,\parallel}$ has roughly a quadratic field dependence over a range of fields up to 5\,T in both samples (neglecting the peak at zero field), whereby the range varies with temperature (see Fig.~\ref{fig:conductivity}).
However, the model cannot explain the saturation of $\sigma_{xx,\parallel}(H)$ in Sample 1 at low temperatures as well as the differences in the low field TMR and LMR between the two samples.
In contrast to Sample 1, the high-field $\sigma_{xx,\parallel}(H)$ in Sample 2 is linear in the accessible field range, congruent with the chiral anomaly model.
An argument that has been brought forward in ref.~\onlinecite{Hirschberger2016} is that the magnetic field-induced changes of the band structure forming the Weyl nodes might cause the deviations in the low-field LMR as well as the dip in the TMR.
Following this argument, magnetic interactions might influence the effective carrier density and relaxation time as well beyond the field range where the dip in the TMR is observed.
Moreover, in case of aligning $\bm{H}$ perpendicular to $\bm{J}$ the positive TMR contribution might superimpose on the otherwise negative MR, unveiled when aligning $\bm{H}$ and $\bm{J}$.
\begin{figure*}[ht!]
	\centering
	\includegraphics[width=17.2cm]{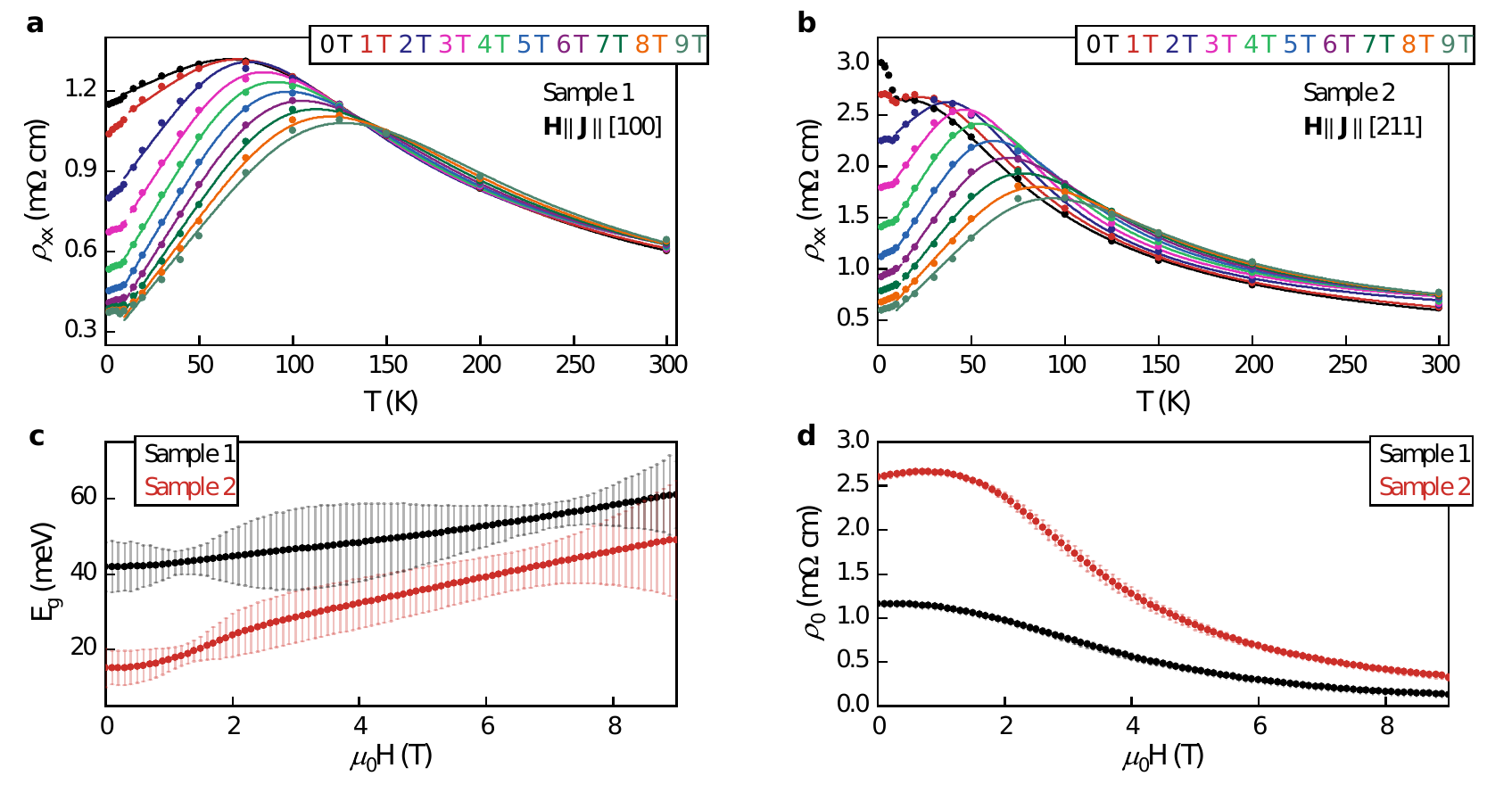}
	\caption{(a,b) Longitudinal magnetoresistance versus $T$ for different magnetic field strenghts for Sample 1 (a) and 2 (b). The solid lines show fits with Eq.~\ref{eq:tempcurve}. (c) Extracted fitting parameter $E_\mathrm{g}$ versus $H$. (d) Fitting parameter $\rho_0$ versus $H$.}
	\label{fig:Fit}
\end{figure*}

A way of tracing the field-induced changes of the electronic properties is to fit the $\rho_{xx,\parallel}(T)$-curves with the phenomenological two-carrier model (introduced in Sec.~\ref{sec:Characterization}, Eq.~\ref{eq:tempcurve}) for constant $H$.
The fits are shown in Figs.~\ref{fig:Fit}a,b and agree well with the data above $T_\mathrm{N}$.
The fit parameters $n_0$, $A$, and $N$ are constant within the error tolerance and only $E_g$ and $\rho_0$ show a smooth evolution with $H$.
The fitting parameters $E_g$ and $\rho_0$ versus $H$ are shown in Figs.~\ref{fig:Fit}c,d.
$E_g$ is increasing with $H$, giving an estimate of the shifting of bands (see Sec.~\ref{sec:Characterization}); whereas $\rho_0$ is decreasing, indicating increased relaxation times with increased fields.
Magnetic-field induced changes of the band structure would also explain the reported\cite{Hirschberger2016} correlation between the strength of the negative LMR and the position of the Fermi level:
the closer the Fermi level is positioned near the band touching point, the bigger the change in the density of states upon shifting of the bands in magnetic fields will be and hence, the bigger the effect on the LMR.
Our samples are exception to this observation, whereas Sample 2 is found to have a LMR $\sim$10\% stronger than Sample 1, with the Fermi level being $\sim$25\,meV closer to the band touching point.
Ultimately, the electron pockets seem to be influenced much more severe, given the relatively constant (within our resolution) SdH frequency of the hole pocket observed in Sample 1 above 5\,T (Fig.~\ref{fig:SdH}f).
It could also be that the field-induced changes in Sample 1 happen mostly at fields below 5\,T, which would be congruent with the bell-shaped LMR curves almost saturating above 6\,T (Fig.~\ref{fig:MR}b).
Another intriguing observation is that both LMR and TMR do not seem to change upon undergoing the antiferromagnetic ordering at 9\,K, showing that ordering of the Gd moments is apparently not crucial for the MR in GdPtBi.
This is surprising, as the opening of the superzone gap leads to a sharp increase of the zero-field $\rho(T)$ (see Fig.~\ref{fig:RvsT}). 
Taking these facts together, a simple spin-disorder scattering mechanism is precluded as an origin of the negative MR, deduced also from the observation that the size of the LMR increases with decreasing carrier density\cite{Hirschberger2016}.
\section{Thermal Magnetotransport}
\label{sec:MTR}
\begin{figure*}[ht!]
	\centering
	\includegraphics[width=17.2cm]{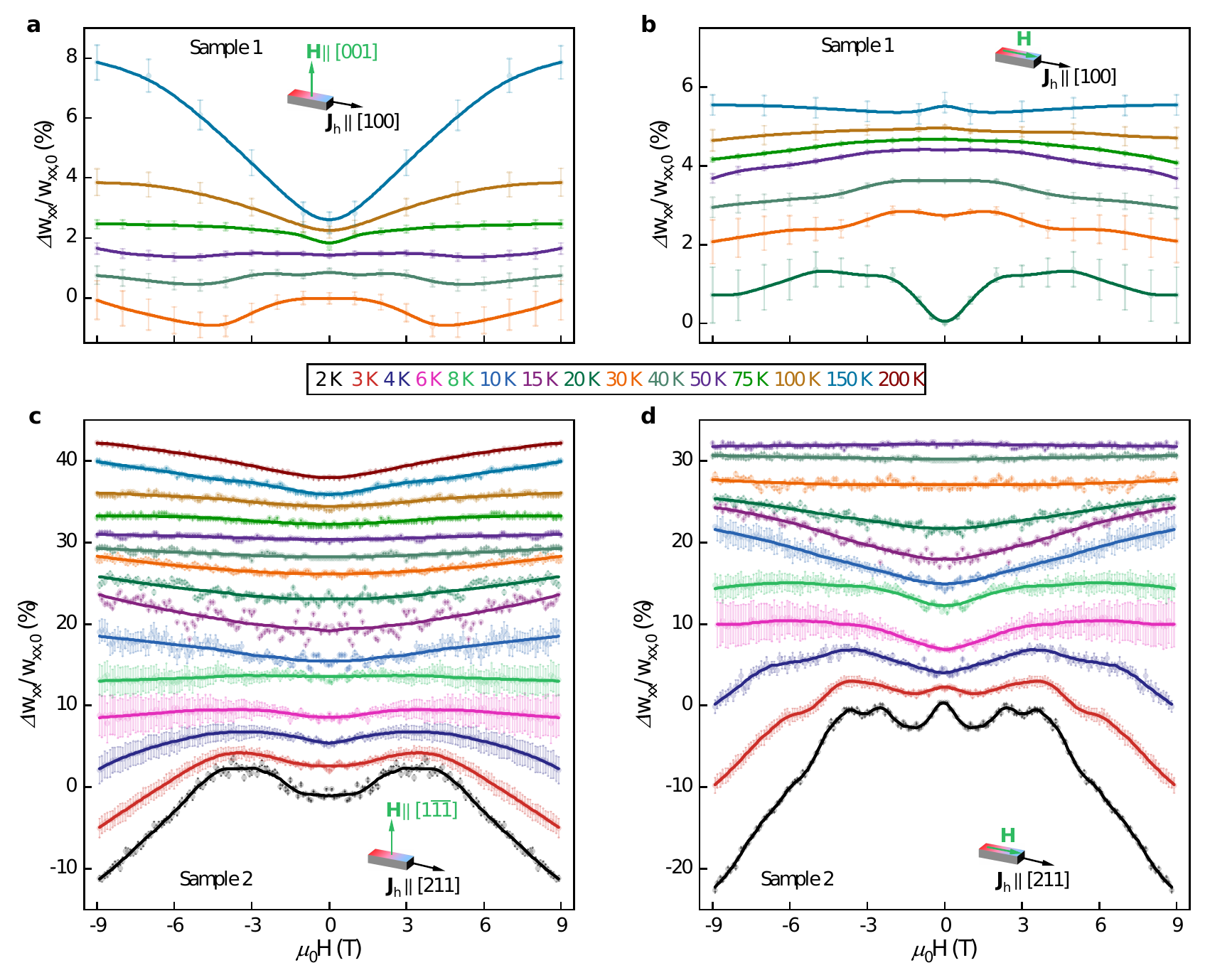}
	\caption{Anisotropic magnetothermal resistivity (MTR) $w_{xx}(H)/w_{xx}(0)$ for Samples 1 and 2. (a,c) Transverse MTR ($\bm{H}\perp\bm{J}_\mathrm{h}$) for different temperatures. (b,d) Longitudinal MTR ($\bm{H}\parallel\bm{J}_\mathrm{h}$). The curves are shifted for better visibility.}
	\label{fig:MTR}
\end{figure*}
In contrast to electrical transport, which can only be carried by electronic excitations, heat transport in GdPtBi can also be carried by phonons and, below $T_\mathrm{N}$, magnons.
Depending on the different scattering mechanisms, phonons and magnons can in principle also exhibit magnetic field-dependencies and cause a non-monotonic MTR.
The MTR of the two samples is shown in Fig.~\ref{fig:MTR}.
During the MTR measurements, the mean temperature $T_\mathrm{mean}=(T_\mathrm{hot}+T_\mathrm{cold})/2$ was found to increase in magnetic field with respect to the base temperature by up to $\sim 0.25\%$ at $\SI{9}{T}$, independent of the $\bm{H}$-orientation.
The systematic error of this heating has been accounted for by multiplying the slope $\mathrm{d}w(T')/\mathrm{d}T$ with $(T_\mathrm{mean}-T')/T'$, yielding the error bars in Fig.~\ref{fig:MTR}.
Above 50\,K, a positive MTR is observed, quadratic in low fields and appearing only for the transverse $\bm{H}$-configuration.
At 150\,K, it leads to an increase of $\sim$4-5\% of the MTR at 9\,T in both samples (Figs.~\ref{fig:MTR}a,c).
Given the estimated contribution of $\kappa_\mathrm{el}$ of 2-5\% (Fig.~\ref{fig:Thermal}), the percentage increase is in the order of the positive electrical TMR (Fig.~\ref{fig:MR}a,d).
The positive transverse MTR therefore likely results from $\kappa_\mathrm{el}$; the thermal two-carrier compensation can be described in a similar way than the electrical TMR.
The electronic thermal conductivity of a two-carrier system in a transverse magnetic field holds\cite{Honig1963}
\begin{equation}
\begin{aligned}
\kappa_{xx}&=\sum\kappa_{1,2}+\frac{T}{D}\Big[\sigma_1\sigma_2(\sigma_1+\sigma_2)(\Delta S_{1,2})^2-\sigma_1\sigma_2\\
\times&\Delta N_{1,2}[(\sigma_1+\sigma_2)\Delta N_{1,2}-2\sigma_1\sigma_2\Delta S_{1,2}(R_1+R_2)]H^2\Big],
\end{aligned}
\label{eq:two-carrierthermal}
\end{equation}
with $\sum\kappa_{1,2}=\kappa_1+\kappa_2$ being the sum of the thermal conductivities of the individual channels.
$\Delta S_{1,2}=S_1-S_2$ and $\Delta N_{1,2}=N_1-N_2$ are the differences of the individual Seebeck and Nernst coefficients, respectively.
Assuming $\kappa_{xy}\ll\kappa_{xx}$ (where $\kappa_{xy}$ is the thermal Hall conductivity), the inverse of Eq.~\ref{eq:two-carrierthermal} describes the two-carrier MTR, resulting in a behavior similar to the two-carrier MR (that is quadratic in low fields and saturating or linear in high fields in case of perfect compensation, respectively).
In contrast to Eq.~\ref{eq:two-carrierMR}, there is a $T$-proportionality, leading to a decrease of the positive MTR upon cooling.
Additionally, $\kappa_\mathrm{el}/\kappa$ strongly decreases with decreasing $T$ (see Fig.~\ref{fig:Thermal}).
Both effects contribute to the rapid vanishing of the positive transverse MTR upon cooling from 150\,K to 50\,K.
In Sample 1, a small negative MTR of $\sim$1\% is observed for the longitudinal configuration below 100\,K and for the transverse configuration below 50\,K.
The shape varies depending on the configuration, from a bell-shaped negative MTR for $\bm{H}\parallel \bm{J}_\mathrm{h}$ to a `W'-shape for $\bm{H}\perp \bm{J}_\mathrm{h}$.
Sample 2 showed no field-dependence in the longitudinal configuration above 50\,K, and only small positive MTR of less than 1\% between 30 and 50\,K for both $\bm{H}$-orientations.
Below 30\,K, $\kappa_\mathrm{el}/\kappa$ drops below 0.1\%, any resolvable field-dependencies of $\kappa$ can therefore no longer originate from the electronic contribution.
As spin-waves only occur in the ordered phase, an abrupt change of $\kappa(T)$ when undergoing the antiferromagnetic phase transition would be expected in case of substantial heat transport by magnons.
Since there was no discernible feature at $T_\mathrm{N}$ (see Fig.~\ref{fig:Thermal}), we assume that the magnon contribution is negligible.
However, even if magnons in GdPtBi do not carry much heat themselves, they can still scatter with phonons and therefore be responsible for the field-dependency of the MTR below $T_\mathrm{N}$.
%
%
%
%
The MTR of Sample 2 was measured down to $T=2\,\mathrm{K}$, showing similarities in the general trend for both $\bm{H}$-configurations.
Below $T_\mathrm{N}$, a `M'-shaped MTR in both transverse and longitudinal MTR (disregarding the additional anomalous features) evolves, which might be explained by phonon scattering from magnons:
If at $H=0$ phonon and magnon dispersion intersect at an energy less than the maximum of the phonon distribution around $4k_\mathrm{B}T$, the thermal resistivity will first increase with rising $H$ and then decrease due to the shift of magnon branches in magnetic field\cite{rareearthbook}.
A $\kappa(H)$-curve of GdPtBi with $\bm{H}\parallel \bm{J}_\mathrm{h}$ at 500\,mK has been reported in ref.~\onlinecite{Hirschberger2017}, showing a similar behavior than the 2\,K-curve of the transverse MTR.
In the longitudinal configuration, the MTR of Sample 2 shows additional features which are not present when $\bm{H\perp\bm{J}_\mathrm{h}}$:
(i) A peak around zero-field evolves upon cooling below 4\,K;
(ii) a dip around 3\,T is observed at 2\,K;
(iii) a deviation from the smooth curve (compared to the transverse MTR) appears at 6\,T below 6\,K.
These features are probably not related to spin-reorientation processes\cite{Coey}, considering the featureless $M(H)$-curve of Sample 2 (Fig.~\ref{fig:Magnetization}).
Such dips might be attributed to resonant scattering with paramagnetic impurities, where the number of scattered phonons depend on the Zeeman splitting at a given $H$, as observed for example in the MTR of Holmium ethylsulphate\cite{rareearthbook,McClintok1967}.
As paramagnetic impurities can generally be present in GdPtBi, especially indicated by in the magnetic susceptibility of Sample 1 (see Sec.~\ref{sec:Characterization}), this might also serve as an explanation for the features in the longitudinal MTR of Sample 2.
In case of an anisotropic $g$-factor of the impurity, this effect might only appear for $\bm{H$} along certain directions.
It might also be that the dispersion of different magnon branches along certain directions leads to field-dependent transitions.
The magnon spectrum of GdPtBi has recently been studied via inelastic neutron scattering\cite{Sukhanov2019}.
Two modes with similar energy at their minima were revealed, but depending on the direction they disperse quite differently.
The MTR in the intermediate temperature range from $T_\mathrm{N}$ to 30\,K, where neither magnons nor electrons are expected to have an influence, might originate from phonon scattering from the paramagnetic Gd-ions as well as impurities.
Scattering from paramagnetic impurities might explain the sample-dependent MTR in that range.
The rich MTR features and the anomalies in the magnetization (Fig.~\ref{fig:Magnetization}) in Sample 1 could therefore be related and might also explain why the MR of Sample 1 exhibits additional features compared to Sample 2.
Furthermore, frustration might also lead to a field-dependent phonon density\cite{Ramirez1994}, however, the moderate frustration of the Gd-ions is found to be relieved already by the next-nearest neighbour-interactions\cite{Sukhanov2019}, GdPtBi thus exhibits only low frustration.
Magnetostriction might have an effect on the MTR as well.
Despite the vanishing crystal field of Gd, magnetoelastic effects in Gd intermetallics can sometimes be of the same order of magnitude as in other rare-earth compounds \cite{magnetostriction}.
Hence, they might as well be significant in GdPtBi.
\section{Summary}
We have studied the magnetization, thermal and electrical magnetotransport in two single-crystalline GdPtBi samples and analyzed the Shubnikov-de Haas oscillations occurring in one of the samples.
Our findings show that the complex, anisotropic magnetotransport in GdPtBi cannot be solely explained within the single-particle Weyl model, but requires to take magnetic interactions of both itinerant charge carriers and phonons with the localized Gd $4f$-moments and possibly also with paramagnetic impurities into account.
These results further stimulate the exploration of the interplay between topological band structure features and many-body interactions.
\begin{acknowledgments}
	Cl.~S. thanks R. Koban for providing technical support and C. Geibel for engaging in clarifying discussions.
	This work was financially supported by the ERC Advanced Grant No. 291472 ’Idea Heusler’ and ERC Advanced Grant No. 742068 ’TOPMAT’.
	Cl.~S. acknowledges financial support by the International Max Planck Research School for Chemistry and Physics of Quantum Materials (IMPRS-CPQM).
	A.~G.~G. thanks J. Cano, B. Bradlyn and R. Ilan for engaging in clarifying discussions.
	A.~G.~G.~acknowledges financial support from the Marie Curie programme under EC Grant agreement No.~653846.
	T.~M.~acknowledges funding by the Deutsche Forschungsgemeinschaft through the Emmy Noether Programme ME 4844/1-1 and through SFB 1143.
\end{acknowledgments}
\end{document}